\documentclass[
]{ceurart}

\usepackage{listings}
\begin{document}

\copyrightyear{2023}
\copyrightclause{Copyright for this paper by its authors.
  Use permitted under Creative Commons License Attribution 4.0
  International (CC BY 4.0).}

\conference{Forum for Information Retrieval Evaluation, December 15-18, 2023, India}

\title{A study of the impact of generative AI-based data augmentation on software metadata classification}

\tnotemark[1]

\author[1]{Tripti Kumari}[%
email=22dr0264@iitism.ac.in,
]
 \cormark[1]
\address[]{Department of Computer Science and Engineering\\
Indian Institute of Technology (ISM) Dhanbad, Jharkhand, 826004, India}
\author[1]{Chakali Sai Charan}[%
email=22mt0348@iitism.ac.in
]
\author[1]{Ayan Das}[%
email=ayandas@iitism.ac.in
]


\cortext[1]{Corresponding author.}

\begin{abstract}
  This paper presents the system submitted by the team from IIT(ISM) Dhanbad in FIRE IRSE 2023 shared task 1 on the automatic usefulness prediction of code-comment pairs as well as the impact of Large Language Model(LLM) generated data on original base data towards an associated source code. We have developed a framework where we train a machine learning-based model using the neural contextual representations of the comments and their corresponding codes to predict the usefulness of code-comments pair and performance analysis with LLM-generated data with base data. In the official assessment, our system achieves a 4\% increase in F1-score from baseline and quality of generated data.
\end{abstract}


\maketitle

\section{Introduction}
\label{sec:intro}
In the rapidly developing world of software development, comments play a crucial role in enhancing code readability and maintainability of the corresponding codes in the code bases~\cite{shinyama2018}. Before executing any software maintenance-related task, or doing any kind of modification and enhancement, developers usually spend a significant amount of time reading and understanding the codes. This process is very time-consuming, particularly, in the case of source codes that implement complex functionalities. So, it is common practice among developers to write comments for code snippets to enhance the comprehensibility of the code. The comments are expected to be helpful in capturing the complete structure and functionality of the codes. This makes commenting one of the most commonly employed documentation methods for software maintenance tasks~\cite{de} on condition that the comments are elaborate and expressive enough to capture the functionality of the programs and that the quality of the comments is maintained throughout the code base.

However, sometimes the comments themselves may be incomplete, inconsistent, and difficult to relate to the source code~\cite{tan}. Such comments may result in a waste of effort in the interpretation of the corresponding code and even may result in a complete misinterpretation of the purpose of the program.
\begin{table}[!ht]
    \centering
    \small
    \caption{Samples of original base data }
   \begin{tabular}{ |l|l|l|l|}
 \hline
 Comment& Surrounding code context  &Label&Explanation\\
 \hline
/*deal with it later*/ & -1.             {/*deal with it later*/1.}&Not Useful& \shortstack[l]{The code does not exist \\ for this comment.\\So, comment is Not Useful}\\
 \hline
 \shortstack[l]{/*switch on*/	}&  \shortstack[l]{-1.f(toggle)\\
 /*switch on*/1.else} & Useful  & \shortstack[l]{The comment correctly\\ describes the code and \\hence Useful}\\
 \hline
\end{tabular}
    \label{tab:dataset}
\end{table}

 Thus understanding the relevance of a comment to a piece of code is crucial before actually using it to understand the purpose of the program. However, given the volume of source code in a standard software project, it is a laborious task to manually verify the usefulness of each comment to their corresponding code. Thus, a system that can automatically predict the usefulness of a comment to its related code snippet may significantly speed up the process of source code analysis. Furthermore, the comment may be rewritten to make it more relevant and informative in case the system predicts the comment to be unuseful.

 Recently, artificial intelligence-based interactive systems, such as ChatGPT~\cite{openai2023} are being widely used to generate texts for different real-time purposes. These systems are also being used by programmers to generate comments for their programs to save time and effort. However, no work has been reported in the literature on the quality of the comments generated by such systems. Given a code-comment pair, these systems may also be used to predict the usefulness of the comment to the corresponding code. However, the accuracy of the predictions of the AI-based systems is not reported in the literature. Thus, it is an open and interesting research area to explore the efficacy of such AI-based systems in automatic comment generation or prediction of the usefulness of a comment for a given code snippet.


The Task-1 of FIRE 2023 IRSE shared task\cite{majumdar2023generative}, mainly focuses on two subtasks. The first subtask is \textit{comment classification}. It involves automatically predicting the usefulness of a given comment to the corresponding source code snippet. It is a binary classification task, that requires us to develop a system, which takes a source code snippet and their associated comment as input. The proposed system automatically classifies whether the comment corresponding to the source code is "Useful" or "Not Useful". The overview paper of IRSE2022 contains information about the shared task in detail\cite{irse2022overview}. We have proposed a system, which takes a code-comment pair as input and generates their representations using a pre-trained neural encoder uses these representations to predict whether the comment is relevant to the code.

The second subtask is \textit{to study the impact of large language models in comments}. In this subtask, the participants are required to augment the base data provided for Subtask 1 with additional data and to carry out a comparative study of the performance of the models trained using the base data and augmented data. The additional data for augmentation is expected to comprise the code-comment pairs obtained from different sources with their usefulness labels predicted using large language models (LLMs)\cite{gao2023enabling}. For this purpose, we manually collect code-comment pairs from different data resources such as GitHub, stack overflow, computer vision, Curl, etc., and then queried the ChatGPT\cite{openai2023} with each code-comment pair to get the usefulness label. We augmented this data with the original seed data and trained some models using different combinations of the additional dataset. We carried out a set of experiments to study the effect of data augmentation on the system performance.

This paper reported the comprehensive explanation of the proposed system submitted to FIRE IRSE2023 for task 1\cite{majumdar2023generative}. We have conducted some experiments and trained the machine learning models, which take the representations of a snippet of source code and their corresponding comments as input. These trained models made predictions about the relevance of the comment to the associated source code with original base data and augmented datasets.

The remaining sections are arranged as follows. Section~\ref{sec:relwork}, presented the related works where we have done some literature surveys on previous work. In this section~\ref{sec:ed}, we have presented a description of different types of LLM-generated datasets and the data made available for the shared task. We have also reported a brief description of data representation and system specification of the system submitted for the shared task. Section~\ref{sec:ra}, presented a comprehensive analysis of the results on different runs with the dataset. In Section~\ref{sec:conclusion}, we have concluded our work on shared tasks.



\section{Related work}
\label{sec:relwork}
We have done some surveys on the usefulness of code-comment pairs as well as the impact of ChatGPT\cite{openai2023} generated comments. We found out some important studies.

Majumdar et al.\cite{majumdar2022}, proposed a survey paper, which is based on the IRSE track (FIRE 2022), and developed solutions for automated evaluation of code comments and classifying comments as useful or not Useful. 
Rahman et al.\cite{rahman2017} did a comparative study on usefulness and developed a RevHelper for automatic usefulness prediction.
Soni et al.\cite{soni2023} developed an automatic text classifier to identify ChatGPT-generated summaries.
Shinyama et al.\cite{shinyama2018}propose a model i.e.C4.5 for code-comments analysis.
Naili et al.\cite{naili2017}developed a generator network with a coverage mechanism. Pre-trained ELMo contextual
embedding was used to generate the highlights of this research paper.
Majumdar et al.\cite{majumdar2020comment} have proposed a COMMENT-MINE semantic search architecture. this architecture is mainly used to extract knowledge based on the design, implementation, and development of software in the form of a knowledge graph.
Majumdar et al.\cite{majumdar2022automated}, developed features to semantically analyze the comments to concepts based on categories of usefulness. They have used Neural networks(NN) to know the usefulness of code comment pairs.
Majumdar et al.\cite{majumdar2022effective} search for contextualized embeddings for code search and classification and developed a system for generating contextualized representations for codes and comments by training ELMo from scratch. 

 

 
\section{Experiment design}
\label{sec:ed}
This section presents a detailed discussion of the proposed system developed for automatically predicting the relevance of code-comment pairs and the experiments carried out on the different combinations of the data sets. In Subsection~\ref{subsec:sd} we present the details of the prediction system. The details for the datasets used for the experiments are reported in Subsection~\ref{subsec:dd}.

\subsection{System description}
\label{subsec:sd}
Our prediction system is a supervised machine-learning-based system that consists of a support vector machine (SVM)~\cite{svm} trained on the distributed representations of the code-comment pairs. It takes the representations of the code-comment pair as input and predicts whether the comment is relevant to the corresponding code snippet.

The distributed representations of code-comment pairs are obtained from a pre-trained ELMO-based model~\cite{emlo}. We have used the ELMO code\footnote{ELMO Code link to generate word embeddings-https://github.com/SMARTKT/WordEmbeddings} provided by the Information Retrieval in Software Engineering (IRSE) team. For a given code comment pair, we separately pass the code and the comment to the ELMO model\cite{emlo} as input as a sequence of tokens. For an input sequence, the ELMO model\cite{emlo} generates 200-dimensional contextual embeddings for each space-separated token in the input sequence. The representation for an input sequence is then obtained by taking the mean of all the token representations. So, the representation of a given input sequence is a 200-dimensional embedding. The 200-dimensional representations of the code and the comment sequences are then concatenated into a joint 400-dimensional representation.

During training, we generate the representations for all the code-comment pairs in the training data and use them to train the support vector machine using the radial basis function (RBF) kernel\cite{thurnhofer2020radial}. During testing, the model saved during the training phase takes the representation of the code-comment pair as input and predicts the usefulness of the comment.

\begin{figure}[htbp]
\centering
\includegraphics[scale=0.5]{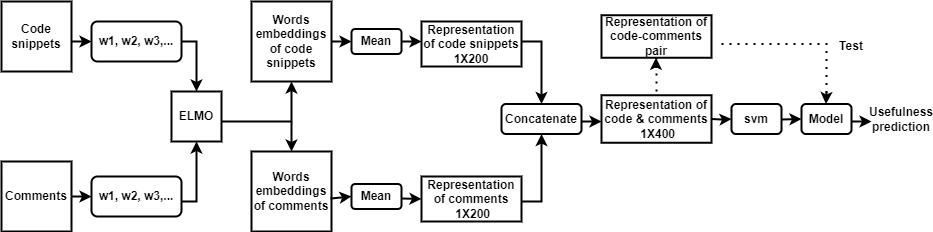}
\caption{Block diagram of proposed model}
\label{fig:model}
\end{figure}

The details of the working of our prediction system is presented in Figure~\ref{fig:model}.




\subsection{Data description}
\label{subsec:dd}
Here we present a description of different combinations of the used datasets for our experiments.

\subsubsection{Original data}
\label{subsubsec:od}
The original data for task 1\cite{majumdar2023generative} was shared by the FIRE IRSE 2023. It contained 11,452 pairs of comments, surrounding code snippets, and their class labels. i.e. if a comment is relevant to the corresponding source code then the corresponding pair is labeled as "Useful" otherwise it is labeled as "Not Useful". A total of 11,452 rows of comments were written in text format and their surrounding source codes. A total of 4,389 code-comment pairs are labeled as "Not Useful" and 7,063 code-comment pairs are labeled as "Useful", which is mentioned in a sample example in Table~\ref{tab:dataset} and Table~\ref{tab:my_label}.


\begin{table}[!ht]
    \centering
    \small
    \caption{Sample of Original base dataset}
   
\begin{tabular}{ |p{5cm}||p{5cm}|p{3cm}|  }
 \hline
 Comment& Surrounding code context  &Label\\
 \hline
 /*upper 8 bit CLASS*/ & -7.if(dot)
-6.host p++;
 & Useful\\
 /*need expand*/ & -1.png set background fixed(png ptr,c;
 & Not Useful\\
  \hline
\end{tabular}
 \label{tab:my_label}
\end{table}

\subsubsection{LLM-generated data}
\label{subsubsec:llmgd}
For Subtask 2, we have manually collected a total of 510 code-comments pairs from different data resources such as GitHub, stack overflow, computer vision, curl, etc., and then query the ChatGPT\cite{openai2023} with each code-comments pair to get the usefulness label. Then we augment this data with the original base data seen in Table~\ref{Tab:dl} and we re-trained the model with the augmented data.

\subsubsection{Extra-generated data}
\label{subsubsec:egd}
We have experimented with another set of data where we have randomly extracted a subset of 250 "Useful" and 250 "Not Useful" code-comment pairs from the original seed data and altered their labels using the following strategy. We converted the "Useful" pairs into "Not Useful" by randomly shuffling the comments and we ensured that at the end of the shuffling none of the codes had their original comments. We labeled such pairs as "Not useful". To convert the "Not Useful" comments to "Useful", we queried the ChatGPT\cite{openai2023} with the code snippets and got the comments synthetically generated. This set of code and synthetically generated comment pairs were labeled as "Useful". Table ~\ref{Tab:dl} gives an explanation of the datasets.

For the sake of convenience, we referred to the original data, LLM-generated data, and extra-generated data as Data1, Data2, and Data3 respectively as shown in Table~\ref{Tab:dl}.
\begin{table}[!ht]
  \centering
  \caption{Different types of data with their size after train test split}
\begin{tabular}{ |p{4cm}||p{3cm}|p{3cm}| }
 \hline
Dataset description & Train dataset size & Test dataset size \\
 \hline
Original data: Data1  & 9162 &2290\\
LLM-generated data: Data2  & 408 &102 \\
Extra-generated data: Data3 & 400 &100\\
\hline
\end{tabular}
\label{Tab:dl}
\end{table}

 

 We have followed the following steps to split the original and the LLM-generated\cite{gao2023enabling} data:
 (i) We have separated out the "Useful" and "Not Useful" code-comment pairs from the data into two groups.
 (ii) We then split each group in an 80:20 ratio.

Thus, the training data comprised a combination of 80\% of the code-comment pairs with "Useful" labels and 80\% of the code-comment pairs with "Not Useful" labels. The selection of the 80\% of the samples in both cases was done randomly. The test data consisted of the remaining 20\% of the samples from both groups.
 
 We followed the same splitting procedure for the  "LLM-generated data"\cite{gao2023enabling} and "Extra generated data" as well. 
 The train and test split size of the dataset is shown in Table~\ref{Tab:dl}.
 \subsection {Combination of all datasets}
 \label{subsec:dsd}
 
 We have created different combinations of datasets by combining different types of datasets. The dataset description is given in this subsection (~\ref{subsubsec:od}, ~\ref{subsubsec:llmgd}, and ~\ref{subsubsec:egd}). Here, we have combined the different data as shown in Table ~\ref{Tab:dcds} with train test split sizes of data. The purpose is to create new datasets to understand the impact of system performance on original data, LLM-generated data\cite{gao2023enabling}, extra-generated data, and different combinations among them shown in Table~\ref{Tab:dcds}. 
\begin{table}[!ht]
\centering
\caption{Different combination of datasets:}
\begin{tabular}{ |p{5cm}||p{3cm}|p{3cm}| }
 \hline
Datasets & Train dataset size & Test dataset size \\
 \hline
Dataset1: Data1 & 9162 &2290\\
Dataset2: Data1+Data2 & 9570 &2392 \\
Dataset3: Data1+Data3  & 9562  &2390\\
Dataset4: Data1+Data2+Data3  & 9970 &2492 \\
\hline
\end{tabular}
\label{Tab:dcds}
\end{table}


\section{Result Analysis}
\label{sec:ra}
In this section, we present a discussion of our results. We performed four different experiments with different combinations of test datasets as shown in Table~\ref{Tab:eap1}, and Table~\ref{Tab:eap2}.



 


\subsection{Run1: Original data (Dataset1)}
\label{subsec:run1}
We performed experiments with the original base data of size 11,452. Original data is split into train and test of sizes 9162 and 2290 in the ratio of 80:20 shown in Table~\ref{Tab:dcds}. We used only the test data of the original base data (Dataset1) for usefulness prediction.

\subsection{Run2: Combination of original data and LLM-generated data (Dataset2)}
\label{subsec:run2}
Our second experiment was carried out with original data and LLM-generated data\cite{gao2023enabling}.
A total of 510 LLM-generated data are split in the ratio of 80:20 into train and test data sizes are 408 and 102 respectively. Now, the total sum of training data and test data of Dataset2 sizes are 9570 and 2392.
To analyze the impact of LLM-generated data\cite{gao2023enabling} on proposed system performance, we augmented test data of original data and LLM-generated data\cite{gao2023enabling} (Dataset2) are shown in Table~\ref{Tab:dcds}.


\subsection{Run3: Original data and extra-generated data (Dataset3)}
\label{subsec:run3}
Our third experiment is with the combination of original base data and extra generated data. A total of 500 extra-generated data are split in the ratio of 80:20 into train and test data sizes are 400 and 100 respectively. Now, the total sum of training data and test data of Dataset3 sizes are 9562 and 2390. 

 \subsection{Run4: Original data, LLM-generated data, and extra-generated data (Dataset4)}
\label{subsec:run4}

 We did one more experiment with the combination of original data, LLM-generated data\cite{gao2023enabling}, and extra-generated data.
 The total sum of training data and test data of Dataset4 sizes are 9970 and 2492. 

\subsection{Result summary}
The overall accuracies corresponding to the experiments carried out for Run1 (Subsection~\ref{subsec:run1}, Run2 (Subsection~\ref{subsec:run2}), Run3 (Subsection~\ref{subsec:run3}) and Run4 (Subsection~\ref{subsec:run4}) are 92.18\%, 92.76\%, 90.696\%, and 92.47\% respectively. The results are summarized in Table~\ref{Tab:eap1}. In Run3 (Dataset3), the accuracy value decreases, and in other Runs (with Dataset1, Dataset2, Dataset3), we are getting almost the same accuracies with slight variation in decimal fractions value.

To evaluate the performance of the system with respect to the "Useful" class, we have used precision, recall, and F1-score as evaluation metrics. The results are summarized in Table~\ref{Tab:eap2}. We have carried out different runs using their corresponding useful class dataset and evaluated the Useful precision, recall, and F1-score.
 In Run1 (Useful dataset size -1465) and Run4 (Useful dataset size- 1578), we are getting the same precision, recall, and F1 score. But, in the case of Run2 (Useful dataset size- 1542), it gets slightly higher recall than other Runs but other evaluation parameters remain the same and in Run2 (Useful dataset size- 1470), all evaluation parameters slightly decrease than other Runs.


\begin{table}
\centering
\caption{Experiment analysis part-1}
\begin{tabular}{ |p{3cm}||p{3cm}|p{3cm}|p{3cm}| }
 \hline
 Experiments& Datasets &Algorithm&Accuracy\\
 \hline
Run1  & Dataset1 &ELMO, SVM&  92.18\\
Run2  & Dataset2 &ELMO, SVM&   92.76\\
Run3  & Dataset3 &ELMO, SVM&  90.696 \\
Run4  & Dataset4 &ELMO, SVM&  92.47\\

  \hline
\end{tabular}
\label{Tab:eap1}
\end{table}

\begin{table}[!ht]
\centering
\caption{Experiment analysis part-2 with "Useful" class}
\begin{tabular}{ |p{2cm}||p{3cm}|p{2.5cm}|p{2cm}|p{2.5cm}|   }
 \hline
 Experiments & Useful dataset size & Useful precision  & Useful recall& Useful F1-score \\
 \hline
Run1   & 1465 & 0.92   &0.96& 0.94\\
Run2   & 1542& 0.92   &0.97& 0.94\\
Run3   & 1470 & 0.89   &0.96& 0.93\\
Run4   & 1578 & 0.92   &0.96& 0.94\\
  \hline
\end{tabular}
\label{Tab:eap2}
\end{table}
\section{Conclusion}
\label{sec:conclusion}
In this paper, we presented our proposed system submitted for participating in task-1 shared by IRSE FIRE 2023. The first task of shared task-1 is to build a system that takes a code-comment pair as input to the encoder, which generates embedding that is passed to the classifier and the classifier classifies whether the comment that corresponds to the code is "Useful" or "Not Useful". The second task is to make predictions on the  augmentation of original seed data and
 LLM-generated data. We have also done impact analysis and model performance with an augmented dataset(original base data and LLM-generated data). All the performance evaluation metrics parameters are mentioned in Table~\ref{Tab:eap1} and Table~\ref{Tab:eap2}. According to the declared result, our system achieves a 4\% increase in F1-score from baseline and quality of data generated.


\bibliography{sample-ceur}

\begin{thebibliography}{17}
\expandafter\ifx\csname natexlab\endcsname\relax\def\natexlab#1{#1}\fi
\providecommand{\url}[1]{\texttt{#1}}
\providecommand{\href}[2]{#2}
\providecommand{\path}[1]{#1}
\providecommand{\DOIprefix}{doi:}
\providecommand{\ArXivprefix}{arXiv:}
\providecommand{\URLprefix}{URL: }
\providecommand{\Pubmedprefix}{pmid:}
\providecommand{\doi}[1]{\href{http://dx.doi.org/#1}{\path{#1}}}
\providecommand{\Pubmed}[1]{\href{pmid:#1}{\path{#1}}}
\providecommand{\bibinfo}[2]{#2}
\ifx\xfnm\relax \def\xfnm[#1]{\unskip,\space#1}\fi
\bibitem[{Shinyama et~al.(2018)Shinyama, Arahori, and Gondow}]{shinyama2018}
\bibinfo{author}{Y.~Shinyama}, \bibinfo{author}{Y.~Arahori}, \bibinfo{author}{K.~Gondow},
\newblock \bibinfo{title}{Analyzing code comments to boost program comprehension},
\newblock in: \bibinfo{booktitle}{2018 25th Asia-Pacific Software Engineering Conference (APSEC)}, \bibinfo{organization}{IEEE}, \bibinfo{year}{2018}, pp. \bibinfo{pages}{325--334}.
\bibitem[{de~Souza et~al.(2005)de~Souza, Anquetil, and de~Oliveira}]{de}
\bibinfo{author}{S.~C.~B. de~Souza}, \bibinfo{author}{N.~Anquetil}, \bibinfo{author}{K.~M. de~Oliveira},
\newblock \bibinfo{title}{A study of the documentation essential to software maintenance},
\newblock \bibinfo{journal}{Association for Computing Machinery}  (\bibinfo{year}{2005}) \bibinfo{pages}{68–75}. \URLprefix \url{https://doi.org/10.1145/1085313.1085331}. \DOIprefix\doi{10.1145/1085313.1085331}.
\bibitem[{Tan et~al.(2007)Tan, Yuan, Krishna, and Zhou}]{tan}
\bibinfo{author}{L.~Tan}, \bibinfo{author}{D.~Yuan}, \bibinfo{author}{G.~Krishna}, \bibinfo{author}{Y.~Zhou},
\newblock \bibinfo{title}{/*icomment: Bugs or bad comments?*/},
\newblock \bibinfo{journal}{SIGOPS Oper. Syst. Rev.} \bibinfo{volume}{41} (\bibinfo{year}{2007}) \bibinfo{pages}{145–158}. \URLprefix \url{https://doi.org/10.1145/1323293.1294276}. \DOIprefix\doi{10.1145/1323293.1294276}.
\bibitem[{OpenAI(2023)}]{openai2023}
\bibinfo{author}{OpenAI}, \bibinfo{title}{Gpt-4 technical report}, \bibinfo{year}{2023}. \href{http://arxiv.org/abs/2303.08774}{{\tt arXiv:2303.08774}}.
\bibitem[{Majumdar et~al.(2023)Majumdar, Paul, Paul, Bandyopadhyay, Dave, Chattopadhyay, Das, Clough, and Majumder}]{majumdar2023generative}
\bibinfo{author}{S.~Majumdar}, \bibinfo{author}{S.~Paul}, \bibinfo{author}{D.~Paul}, \bibinfo{author}{A.~Bandyopadhyay}, \bibinfo{author}{B.~Dave}, \bibinfo{author}{S.~Chattopadhyay}, \bibinfo{author}{P.~P. Das}, \bibinfo{author}{P.~D. Clough}, \bibinfo{author}{P.~Majumder},
\newblock \bibinfo{title}{Generative ai for software metadata: Overview of the information retrieval in software engineering track at fire 2023},
\newblock in: \bibinfo{booktitle}{Forum for Information Retrieval Evaluation, ACM}, \bibinfo{year}{2023}.
\bibitem[{Majumdar et~al.(2022)Majumdar, Bandyopadhyay, Das, D~Clough, Chattopadhyay, and Majumder}]{irse2022overview}
\bibinfo{author}{S.~Majumdar}, \bibinfo{author}{A.~Bandyopadhyay}, \bibinfo{author}{P.~P. Das}, \bibinfo{author}{P.~D~Clough}, \bibinfo{author}{S.~Chattopadhyay}, \bibinfo{author}{P.~Majumder},
\newblock \bibinfo{title}{{Overview of the IRSE track at FIRE 2022: Information Retrieval in Software Engineering}},
\newblock in: \bibinfo{booktitle}{Forum for Information Retrieval Evaluation}, \bibinfo{publisher}{ACM}, \bibinfo{year}{2022}.
\bibitem[{Gao et~al.(2023)Gao, Yen, Yu, and Chen}]{gao2023enabling}
\bibinfo{author}{T.~Gao}, \bibinfo{author}{H.~Yen}, \bibinfo{author}{J.~Yu}, \bibinfo{author}{D.~Chen},
\newblock \bibinfo{title}{Enabling large language models to generate text with citations},
\newblock \bibinfo{journal}{arXiv preprint arXiv:2305.14627}  (\bibinfo{year}{2023}).
\bibitem[{Majumdar et~al.(2022)Majumdar, Bandyopadhyay, Das, Clough, Chattopadhyay, and Majumder}]{majumdar2022}
\bibinfo{author}{S.~Majumdar}, \bibinfo{author}{A.~Bandyopadhyay}, \bibinfo{author}{P.~P. Das}, \bibinfo{author}{P.~Clough}, \bibinfo{author}{S.~Chattopadhyay}, \bibinfo{author}{P.~Majumder},
\newblock \bibinfo{title}{Can we predict useful comments in source codes?-analysis of findings from information retrieval in software engineering track@ fire 2022},
\newblock in: \bibinfo{booktitle}{Proceedings of the 14th Annual Meeting of the Forum for Information Retrieval Evaluation}, \bibinfo{year}{2022}, pp. \bibinfo{pages}{15--17}.
\bibitem[{Rahman et~al.(2017)Rahman, Roy, and Kula}]{rahman2017}
\bibinfo{author}{M.~M. Rahman}, \bibinfo{author}{C.~K. Roy}, \bibinfo{author}{R.~G. Kula},
\newblock \bibinfo{title}{Predicting usefulness of code review comments using textual features and developer experience},
\newblock in: \bibinfo{booktitle}{2017 IEEE/ACM 14th International Conference on Mining Software Repositories (MSR)}, \bibinfo{organization}{IEEE}, \bibinfo{year}{2017}, pp. \bibinfo{pages}{215--226}.
\bibitem[{Soni and Wade(2023)}]{soni2023}
\bibinfo{author}{M.~Soni}, \bibinfo{author}{V.~Wade}, \bibinfo{title}{Comparing abstractive summaries generated by chatgpt to real summaries through blinded reviewers and text classification algorithms}, \bibinfo{year}{2023}. \href{http://arxiv.org/abs/2303.17650}{{\tt arXiv:2303.17650}}.
\bibitem[{Naili et~al.(2017)Naili, Chaibi, and Ghezala}]{naili2017}
\bibinfo{author}{M.~Naili}, \bibinfo{author}{A.~H. Chaibi}, \bibinfo{author}{H.~H.~B. Ghezala},
\newblock \bibinfo{title}{Comparative study of word embedding methods in topic segmentation},
\newblock \bibinfo{journal}{Procedia computer science} \bibinfo{volume}{112} (\bibinfo{year}{2017}) \bibinfo{pages}{340--349}.
\bibitem[{Majumdar et~al.(2020)Majumdar, Papdeja, Das, and Ghosh}]{majumdar2020comment}
\bibinfo{author}{S.~Majumdar}, \bibinfo{author}{S.~Papdeja}, \bibinfo{author}{P.~P. Das}, \bibinfo{author}{S.~K. Ghosh},
\newblock \bibinfo{title}{Comment-mine—a semantic search approach to program comprehension from code comments},
\newblock \bibinfo{journal}{Advanced Computing and Systems for Security: Volume Twelve}  (\bibinfo{year}{2020}) \bibinfo{pages}{29--42}.
\bibitem[{Majumdar et~al.(2022{\natexlab{a}})Majumdar, Bansal, Das, Clough, Datta, and Ghosh}]{majumdar2022automated}
\bibinfo{author}{S.~Majumdar}, \bibinfo{author}{A.~Bansal}, \bibinfo{author}{P.~P. Das}, \bibinfo{author}{P.~D. Clough}, \bibinfo{author}{K.~Datta}, \bibinfo{author}{S.~K. Ghosh},
\newblock \bibinfo{title}{Automated evaluation of comments to aid software maintenance},
\newblock \bibinfo{journal}{Journal of Software: Evolution and Process} \bibinfo{volume}{34} (\bibinfo{year}{2022}{\natexlab{a}}) \bibinfo{pages}{e2463}.
\bibitem[{Majumdar et~al.(2022{\natexlab{b}})Majumdar, Varshney, Das, Clough, and Chattopadhyay}]{majumdar2022effective}
\bibinfo{author}{S.~Majumdar}, \bibinfo{author}{A.~Varshney}, \bibinfo{author}{P.~P. Das}, \bibinfo{author}{P.~D. Clough}, \bibinfo{author}{S.~Chattopadhyay},
\newblock \bibinfo{title}{An effective low-dimensional software code representation using bert and elmo},
\newblock in: \bibinfo{booktitle}{2022 IEEE 22nd International Conference on Software Quality, Reliability and Security (QRS)}, \bibinfo{organization}{IEEE}, \bibinfo{year}{2022}{\natexlab{b}}, pp. \bibinfo{pages}{763--774}.
\bibitem[{Cortes and Vapnik(1995)}]{svm}
\bibinfo{author}{C.~Cortes}, \bibinfo{author}{V.~Vapnik},
\newblock \bibinfo{title}{Support-vector networks},
\newblock \bibinfo{journal}{Machine learning} \bibinfo{volume}{20} (\bibinfo{year}{1995}) \bibinfo{pages}{273--297}.
\bibitem[{Peters et~al.(2018)Peters, Neumann, Iyyer, Gardner, Clark, Lee, and Zettlemoyer}]{emlo}
\bibinfo{author}{M.~E. Peters}, \bibinfo{author}{M.~Neumann}, \bibinfo{author}{M.~Iyyer}, \bibinfo{author}{M.~Gardner}, \bibinfo{author}{C.~Clark}, \bibinfo{author}{K.~Lee}, \bibinfo{author}{L.~Zettlemoyer},
\newblock \bibinfo{title}{Deep contextualized word representations},
\newblock \bibinfo{journal}{Association for Computational Linguistics}  (\bibinfo{year}{2018}) \bibinfo{pages}{2227--2237}. \URLprefix \url{https://aclanthology.org/N18-1202}. \DOIprefix\doi{10.18653/v1/N18-1202}.
\bibitem[{Thurnhofer-Hemsi et~al.(2020)Thurnhofer-Hemsi, L{\'o}pez-Rubio, Molina-Cabello, and Najarian}]{thurnhofer2020radial}
\bibinfo{author}{K.~Thurnhofer-Hemsi}, \bibinfo{author}{E.~L{\'o}pez-Rubio}, \bibinfo{author}{M.~A. Molina-Cabello}, \bibinfo{author}{K.~Najarian},
\newblock \bibinfo{title}{Radial basis function kernel optimization for support vector machine classifiers},
\newblock \bibinfo{journal}{arXiv preprint arXiv:2007.08233}  (\bibinfo{year}{2020}).

\end{thebibliography}

\appendix

\end{document}